\documentclass[aps,prb,twocolumn,superscriptaddress,showpacs]{revtex4}

\usepackage{graphicx}
\usepackage{dcolumn}
\usepackage{bm}

\begin{document}
\title{An irreversible magnetic-field dependence of low-temperature heat transport
of spin-ice compound Dy$_2$Ti$_2$O$_7$ in a [111] field}

\author{C. Fan}
\affiliation{Hefei National Laboratory for Physical Sciences at
Microscale, University of Science and Technology of China, Hefei,
Anhui 230026, People's Republic of China}

\author{Z. Y. Zhao}
\affiliation{Hefei National Laboratory for Physical Sciences at
Microscale, University of Science and Technology of China, Hefei,
Anhui 230026, People's Republic of China}

\author{H. D. Zhou}
\affiliation{Department of Physics and Astronomy, University of
Tennessee, Knoxville, Tennessee 37996-1200, USA}
\affiliation{National High Magnetic Field Laboratory, Florida
State University, Tallahassee, Florida 32306-4005, USA}

\author{X. M. Wang}
\affiliation{Hefei National Laboratory for Physical Sciences at
Microscale, University of Science and Technology of China, Hefei,
Anhui 230026, People's Republic of China}

\author{Q. J. Li}
\affiliation{Hefei National Laboratory for Physical Sciences at
Microscale, University of Science and Technology of China, Hefei,
Anhui 230026, People's Republic of China}

\author{F. B. Zhang}
\affiliation{Hefei National Laboratory for Physical Sciences at
Microscale, University of Science and Technology of China, Hefei,
Anhui 230026, People's Republic of China}

\author{X. Zhao}
\affiliation{School of Physical Sciences, University of Science
and Technology of China, Hefei, Anhui 230026, People's Republic of
China}

\author{X. F. Sun}
\email{xfsun@ustc.edu.cn} \affiliation{Hefei National Laboratory
for Physical Sciences at Microscale, University of Science and
Technology of China, Hefei, Anhui 230026, People's Republic of
China}

\date{\today}

\begin{abstract}

We study the low-temperature thermal conductivity ($\kappa$) of
Dy$_2$Ti$_2$O$_7$ along and perpendicular to the (111) plane and
under the magnetic field along the [111] direction. Besides the
step-like decreases of $\kappa$ at the field-induced transitions
from the spin-ice state to the kagom\'e-ice state and then to the
polarized state, an abnormal phenomenon is that the $\kappa(H)$
isotherms show a clear irreversibility at very low temperatures
upon sweeping magnetic field up and down. This phenomenon
surprisingly has no correspondence with the well-known
magnetization hysteresis. Possible origins for this
irreversibility are discussed; in particular, a pinning effect of
magnetic monopoles in spin ice compound by the weak disorders is
proposed.

\end{abstract}


\maketitle

\section{INTRODUCTION}

The isolated magnetic charges or the magnetic monopoles, in
contrast to their electrical counterparts, have not been observed
in nature. A recent discovery was that the magnetic monopoles
could emerge from the collective excitations of spin ice.
\cite{monopole-1, monopole-2, monopole-3, monopole-4, monopole-5,
monopole-6, monopole-7, monopole-8, monopole-9, monopole-10}
Although this kind of magnetic-monopole excitations is not
completely the same as the free elementary particles, this finding
is interesting and very important in the sense that it was the
first time to signify the magnetic monopoles in real space.

\begin{figure}
\includegraphics[clip,width=8.5cm]{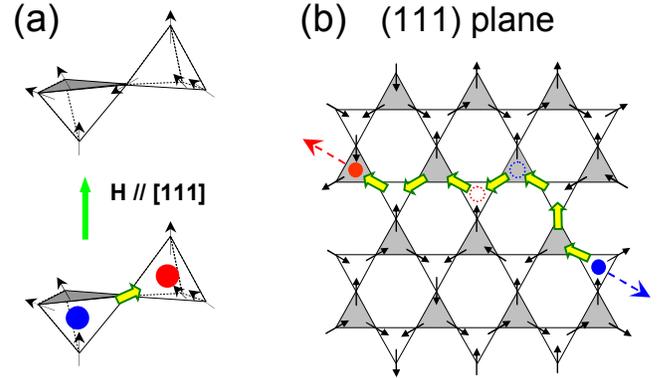}
\caption{(Color online)  (a) ``2-in, 2-out" spin configuration of
the spin-ice ground state in Dy$_2$Ti$_2$O$_7$. The spin-easy axis
is along the three-fold axis or the [111] direction. Two
neighboring tetrahedra are shown and the arrows stand for
Dy$^{3+}$ spins on the vertexes. When one spin is flipped,
indicated by the open arrow, by thermal fluctuations or external
magnetic field, a pair of magnetic monopoles are formed and shown
by the red and blue dots. (b) In magnetic field perpendicular to
the (111) plane, the magnetic monopoles can move in the plane by
successive spin flipping. The dashed red and blue arrows,
accompanied with the dashed open circles, show the routines of
positive and negative monopoles.}
\end{figure}

Dy$_2$Ti$_2$O$_7$ (DTO), as a spin ice material, has already
become the focus of condensed matter physics \cite{monopole-1} and
received a lot of attentions during past ten years due to its
fantastic physical properties.\cite{spin ice-1, spin ice-2, spin
ice-3, spin ice-4, spin ice-5, magnetic transition-1, magnetic
transition-2, magnetic transition-3, magnetic transition-4} DTO
belongs to the family of pyrochlore titanates with the Dy$^{3+}$
spins locating at the vertexes of tetrahedra, which consist of
triangular and kagom\'e planes stacked alternatingly along the
[111] direction. At low temperatures, the Dy$^{3+}$ spins are
Ising anisotropic and form a macroscopically degenerated ``2-in,
2-out'' spin-ice ground state, which is attributed to the
frustration effect between the nearest-neighboring
antiferromagnetic interaction and the ferromagnetic dipolar
interaction.\cite{spin ice-2, spin ice-3, spin ice-5} Once the
flipping of a spin occurs, a local ``3-in, 1-out'' or ``3-out,
1-in'' spin configuration forms, which is equivalent to yielding
two opposite magnetic monopoles in the adjacent
tetrahedra,\cite{magnetic transition-1, magnetic transition-2,
magnetic transition-3} as shown in Fig. 1(a). The monopole pairs
can be separated and diffuse freely in the spin-ice state, which
can be viewed as a ``vacuum'' free of charges by continuous spin
flip and behaves a random walk process in zero
field.\cite{monopole-1, monopole-4} A particular case of interest
is with the magnetic field along the three-fold axis or the [111]
direction. With increasing field, there are two successive
transitions from the spin-ice state to the kagom\'e-ice state and
then to the fully-polarized state.\cite{magnetic transition-1,
magnetic transition-2, magnetic transition-3, magnetic
transition-4} In the kagom\'e-ice state, the magnetic field pins
the spins on the triangular planes along the field direction.
However, the ice rule can still be satisfied in low magnetic
field. As a result, a reduced degeneracy is expected for the
kagom\'e plane and the dimensionality of the spin system is
reduced. Therefore, the magnetic monopoles can diffuse only in the
kagom\'e plane,\cite{monopole-3} as shown in Fig. 1(b). However,
it should be noted that the experimental investigations on probing
the monopole excitations have not yet arrived perfect
consistency.\cite{muonSR-1, muonSR-2} Studying spin-ice materials
by using more different techniques is probably an effective way.

Low-temperature heat transport is a powerful tool to probe the
properties of elementary excitations\cite{Berman, Brenig, Hess,
Sologubenko1, Yamashita, Taillefer, Sun_LSCO, Sun_YBCO} and the
magnetic-field-induced magnetic transitions\cite{CuGeO3,
Sologubenko2, Sologubenko3, Sun_DTN, Wang_HMO, Ke_BMO, Zhao_NCO,
Zhao_BCVO, Wang_TMO}. In principle, the magnetic monopoles, as the
elementary excitations in the spin-ice compounds, can also
contribute to the heat transport by acting as either heat carriers
or phonon scatterers. In this work, we study the low-$T$ thermal
conductivity ($\kappa$) of DTO single crystals down to 0.3 K with
heat current parallel and perpendicular to the (111) plane and in
the magnetic field along the [111] axis. It is found that the
phonon heat transport shows step-like decreases at the magnetic
phase transitions, which is nearly isotropic on the direction of
heat current and is mainly due to the magnetic scattering on
phonons. Besides, a remarkable hysteresis of $\kappa(H)$ isotherm
appears at very low temperatures. The peculiarity is that this
irreversibility has no direct correspondence with the well-known
magnetization irreversibility. The possible mechanisms are
discussed.

\section{EXPERIMENTS}

High-quality DTO single crystals were grown using the
floating-zone technique. The thermal conductivity at low
temperatures down to 0.3 K was measured using a conventional
steady-state technique.\cite{Sun_DTN, Wang_HMO, Zhao_NCO,
Zhao_BCVO} In this work, we measured thermal conductivities along
and perpendicular to the (111) plane, named as
$\kappa_{\parallel}$ and $\kappa_{\perp}$, by using two samples
with sizes of $5.3 \times 0.74 \times 0.16$ mm$^3$ and $3.1 \times
0.71 \times 0.15$ mm$^3$, respectively. Note that the heat
currents have to be along the longest dimension, while the
magnetic field is always perpendicular to the (111) plane and is
along the shortest and the longest dimension for
$\kappa_{\parallel}$ and $\kappa_{\perp}$ samples, respectively.
So the demagnetization effect is unavoidably different for these
two samples. The typical misalignment of the magnetic field is
less than 2$^\circ$.

It is necessary to point out that in this work the magnetic-field
dependencies of $\kappa$ were measured in a static field mode.
That is, the measurements were done by the following steps: (i)
change field slowly to a particular value and keep it stable; (2)
after the sample temperature is stabilized, apply a heat power at
the free end of sample; (iii) wait {\it long enough} time until
the temperature gradient on the sample (judged from the time
dependencies of two RuO$_2$ thermometers on the samples) is
completely stabilized; (iv) record the temperatures of two
thermometers and obtain the temperature gradient. The purpose of
this operation is to avoid any misleading data. It is known that
DTO has very slow spin dynamics at low
temperatures,\cite{Matsuhira} which may cause some relaxation
phenomenon in the temperature gradient measurements, if the
magnetic excitations are involved in the heat transport behaviors.
Actually, in such case, a thermal conductivity measurement in the
sweeping-field mode could be incorrect because it is impossible to
get a steady heat flow.

\section{RESULTS AND DISCUSSION}

\subsection{Heat carriers}

\begin{figure}
\includegraphics[clip,width=8.5cm]{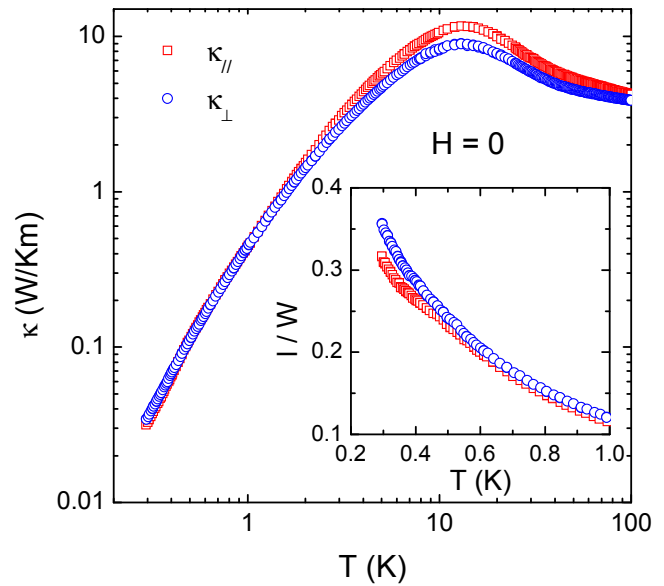}
\caption{(Color online) Temperature dependencies of the thermal
conductivities of Dy$_2$Ti$_2$O$_7$ in zero field for heat current
along the (111) plane or perpendicular to it. The inset shows the
temperature dependencies of the phonon mean free path $l$ divided
by the averaged sample width $W$ for two sets of data.}
\end{figure}

Figure 2 shows the temperature dependencies of
$\kappa_{\parallel}$ and $\kappa_{\perp}$ in zero field. Both
$\kappa_{\parallel}$ and $\kappa_{\perp}$ exhibit clear phonon
peaks at $\sim$ 15 K and a rough $T^{2.7}$ dependence at subKelvin
temperatures, which is however weaker than the standard $T^3$
behavior of phonon thermal conductivity at the boundary scattering
limit.\cite{Berman} Similar result has been obtained for heat
current along the [110] direction, and the magnetic scattering on
phonons was discussed to be important at temperatures below 10
K.\cite{Klemke} It is possible to estimate the mean free path of
phonons at low temperatures and to judge whether the phonons are
free from microscopic scattering at subKelvin temperatures. The
phononic thermal conductivity can be expressed by the kinetic
formula $\kappa_{ph} = \frac{1}{3}Cv_pl$,\cite{Berman} where $C =
\beta T^3$ is the phonon specific heat at low temperatures, $v_p$
is the average velocity and $l$ is the mean free path of phonons.
Using the $\beta$ value obtained from specific-heat measurements
(not shown here), the phonon velocity can be calculated and then
the mean free path is obtained from the $\kappa$.\cite{Zhao_GFO,
Zhao_NCO} The inset to Fig. 2 shows the ratios $l/W$ for the two
samples, where $W$ is the averaged sample width.\cite{Zhao_GFO,
Zhao_NCO} It is clear that both ratios increase quickly at very
low temperatures and are expected to approach 1 below 0.3 K. This
estimation tells us that it is not likely that the magnetic
excitations in DTO (for examples, magnetic monopoles) can make a
sizable contribution to transporting heat, since the experimental
$\kappa$ is even smaller than the phonon term in a boundary
scattering limit. It should be pointed out that the
magnetic-monopole excitations were believed to be negligible at
very low temperatures, because the energy barrier for flipping a
spin or creating a pair of magnetic monopoles is about 4.4
K.\cite{LDC}

\subsection{$\kappa(H)$ and field-induced magnetic transitions}

\begin{figure}
\includegraphics[clip,width=8.5cm]{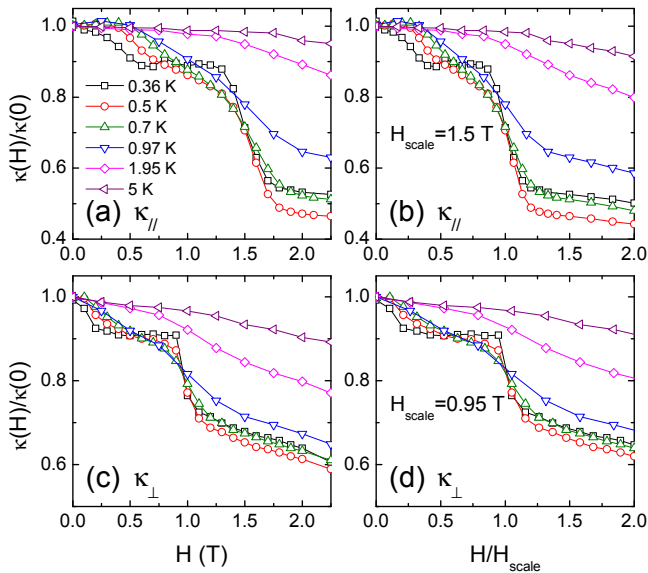}
\caption{(Color online) Magnetic-field dependencies of the
$\kappa_{\parallel}$ and $\kappa_{\perp}$ for field along the
[111] direction. The data are taken with decreasing field from
high-field limit (14 T). Panels (a) and (c) show the raw data,
while (b) and (d) show the data with magnetic fields re-scaled by
1.5 and 0.95 T, respectively, considering the demagnetization
effect. Note that the data after this re-scaling are essentially
isotropic.}
\end{figure}

Figures 3(a) and 3(c) show the $\kappa(H)$ isotherms for
$\kappa_{\parallel}$ and $\kappa_{\perp}$, respectively, in
magnetic field perpendicular to the (111) plane. Note that these
data are taken with sweeping field down from 14 T. One can easily
notice that at very low temperatures $\kappa_{\parallel}(H)$ and
$\kappa_{\perp}(H)$ display two anomalies at low magnetic fields.
For example, at 0.36 K the $\kappa_{\perp}(H)$ curve shows two
clear step-like transitions at about 0.25 and 0.9 T, which are in
good correspondence with the critical fields of the subsequential
transitions from the low-field spin-ice phase to the kagom\'e-ice
phase then to the saturated state (``3-in, 1-out" or ``3-out,
1-in").\cite{magnetic transition-1,magnetic transition-2,magnetic
transition-3} In present work, we defined these two transition
fields as $H_{c1}$ and $H_{c2}$. The sharpness of the transitions
is quickly smeared out upon increasing temperature. Similar
results are observed in the low-$T$ $\kappa_{\parallel}(H)$
isotherms. It is clearly seen that the main difference between the
$\kappa_{\parallel}$ and $\kappa_{\perp}$ is that the critical
fields of the step-like transitions are somewhat different.
However, if the horizontal axis of the $\kappa_{\parallel}(H)$ and
$\kappa_{\perp}(H)$ curves are re-scaled by some particular
values, as shown in Figs. 3(b) and 3(d), one can obtain the
essentially same behavior of $\kappa_{\parallel}$ and
$\kappa_{\perp}$. This means that the difference between the
experimental data of $\kappa_{\parallel}$ and $\kappa_{\perp}$ is
simply due to the demagnetization effect.\cite{monopole-10}
Therefore, the heat transport behaviors shown in Fig. 3 actually
indicate a nearly isotropic behavior and a close relationship
between the heat transport and the magnetic phase transitions.

The drastic change of heat transport at the magnetic phase
transition has been widely observed in many kinds of magnetic
materials and is closely associated with the evolution of magnetic
excitations.\cite{CuGeO3, Sologubenko2, Sologubenko3, Sun_DTN,
Wang_HMO, Ke_BMO, Zhao_NCO, Zhao_BCVO, Wang_TMO} However, the
particular features of $\kappa(H)$ at the magnetic transitions can
be significantly different from each other, which depends on both
the natures of magnetic transitions and the roles of magnetic
excitations in the heat transport (as heat carriers or phonon
scatterers). In principle, the magnetic monopoles as a kind of
excitations in the spin-ice and kagom\'e-ice states can either
transport heat or scatter phonons. However, it is easy to know
that the magnetic monopoles are not likely to act as heat carriers
for two reasons. First, as mentioned above, in zero field the
monopole excitations are actually very difficult to emerge at
subKelvin temperatures. Second, a low field along the [111]
direction can significantly enhance the excitations of magnetic
monopoles, whereas the thermal conductivity does not show any
increase. Then, can the step-like decreases of $\kappa$ at two
transitions be attributed to phonon scattering by the sudden
increases of magnetic monopoles? There is no doubt that the
magnetic monopoles can effectively scatter phonons by absorbing
phonon energy and flipping spins, and suppress the phonon heat
conductivity. However, above $H_{c2}$ ($\sim$ 1--1.5 T), the
magnetic monopoles are so populated that they fully occupy the
kagom\'e network with the monopole pairs sitting on the nearest
sites, in which case their ability of scattering phonons must be
negligible. Therefore, even if the step-like decrease of
$\kappa(H)$ at $H_{c1}$ can be explained by the enhanced phonon
scattering by magnetic monopoles, the one at $H_{c2}$ cannot be
related to the magnetic monopoles.

It is known that, due to the strong Ising anisotropy of Dy$^{3+}$
spins, the magnetic monopoles connected by the Dirac strings are
the elementary magnetic excitations of the spin ice state. The
$\kappa(H)$ behaviors shown in Fig. 3, however, indicate that
there may be other magnetic excitations playing an important role
in the step-like decreases of $\kappa$ at two magnetic
transitions. In this regard, there has been no obvious evidence
that DTO could have other magnetic excitations than the magnetic
monopoles, like the spin fluctuations caused by non-Ising terms in
the Hamiltonian.\cite{muonSR-2, Quemerais} Nevertheless, some
drastic field-induced changes related to crystal lattice and
phonons have been indeed found. In a recent work, the ultrasound
measurements also indicated sharp anomalies of the sound velocity
and sound attenuation at $H_{c2}$, which may share a common origin
with the $\kappa(H)$ transitions.\cite{monopole-10} It is also
notable that a local maximum of entropy at $H_{c2}$ may also
indicate a strong spin fluctuations that can strongly scatter
phonons.\cite{Entropy}

Another important impact on the field dependence of $\kappa$ that
should be taken into account is the paramagnetic scattering effect
on phonons.\cite{Berman, Sun_GBCO, Li_NGSO} It is known that the
Dy$^{3+}$ ions have a degenerate doublet of the lowest
crystal-field level, which was evidenced as a simple two-level
Schottky anomaly of the low-$T$ specific heat in magnetic
fields.\cite{Hiroi, Higashinaka} Therefore, a Zeeman-effect
splitting of the doublet can produce resonant scattering on
phonons and give a low-field suppression of thermal conductivity,
as some other magnetic materials have shown.\cite{Sun_GBCO,
Li_NGSO} This can also explain the continuous decrease of $\kappa$
at relatively high temperatures (above $\sim$ 1 K), where two
magnetic transitions are gone.

\subsection{Irreversibility of $\kappa(H)$ curves}

\begin{figure}
\includegraphics[clip,width=8.5cm]{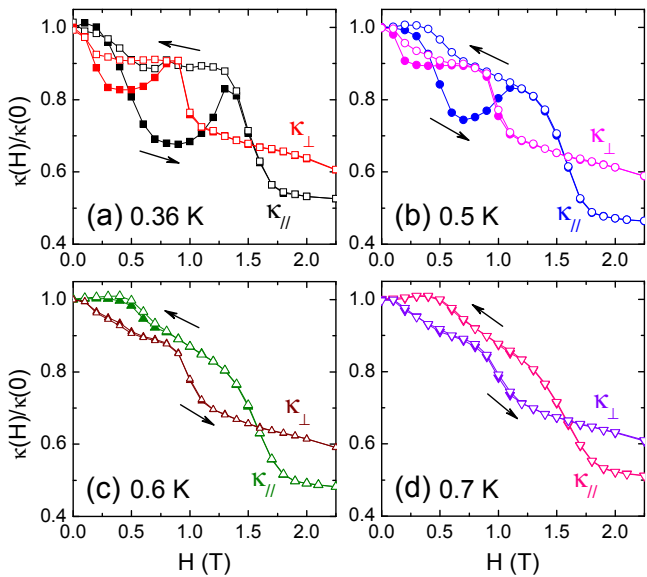}
\caption{(Color online) $\kappa_{\parallel}(H)$ and
$\kappa_{\perp}(H)$ loops with magnetic field sweeping up and down
(perpendicular to the (111) plane). The data shown with solid
symbols are measured in ascending field after the sample is cooled
in zero field, while the open symbols show the data with
descending field, as indicated by arrows. The demagnetization
effects, which are the same as those in Fig. 3, are not taken into
account in these plots.}
\end{figure}

An unusual phenomenon in DTO is that, as shown in Fig. 4, the
$\kappa_{\parallel}(H)$ and $\kappa_{\perp}(H)$ isotherms show an
irreversibility at very low temperatures. For example, after
cooling the sample to 0.36 K in zero field, both
$\kappa_{\parallel}$ and $\kappa_{\perp}$ are firstly measured
with increasing field, for which they show a valley-and-peak
feature. Then, upon decreasing field from above 2 T, the
magnitudes of $\kappa_{\parallel}$ and $\kappa_{\perp}$ deviate
from those of the field-increasing curves below $\sim$ 1.5 and 1
T, respectively. Actually, the differences between field-up and
field-down curves are so large that the valley-and-peak feature of
$\kappa(H)$ does not show up at all for decreasing field. Upon
increasing temperature, the irreversibility is quickly diminished
and disappears above 0.7 K. The peculiarity of this transport
irreversibility is that it has no direct correspondence to that of
the low-$T$ magnetization curves $M(H)$,\cite{magnetic
transition-1} which feature a pronounced hysteresis below 0.5 T
and a small one around 0.9 T that were ascribed to the slow
dynamics of the spin ice and the first-order phase transition,
respectively. Furthermore, some higher-field hysteresis of $M(H)$
was observable only below 0.36 K.\cite{magnetic transition-1}

Although the irreversibility of thermal conductivity can sometimes
be observed at the field-induced first-order phase transitions in
many materials, they are usually very small and presented in a
narrow vicinity of the critical fields.\cite{CuGeO3} One exception
for showing large hysteretic heat transport is the high-$T_c$
superconductors in the mixed state,\cite{Vortex} for which the
hysteretic $\kappa(H)$ behaviors are accompanied with the
magnetization irreversibility caused by the vortex pinning. In
other words, the difference of $\kappa(H)$ between the field-up
curve and the field-down one is directly related to the magnetism,
which is reasonable because the density and distribution of the
vortices determine the strength of vortex-electron scattering and
the electron heat transport.\cite{Franz} In contrast, the
irreversible $\kappa(H)$ behaviors of DTO can exist even in the
field region where the magnetization shows a nearly reversible
plateau.\cite{magnetic transition-1} In addition, the ultrasound
properties also displayed a broad hysteresis, but only at
temperatures as low as 0.29 K.\cite{monopole-10} Since the heat
transport measurement is performed in a static-field mode, in
contrast to the recent interesting nonstationary magnetization and
ultrasound measurements,\cite{monopole-7, monopole-10} it may
reveal different physics from other techniques.

It is necessary to point out that the step-like decreases in the
low-$T$ $\kappa(H)$ isotherms in some sense are not extraordinary,
since similar anomalies have been observed in other magnetic
materials\cite{CuGeO3, Zhao_BCVO} and are not difficult to
understand. Although the hysteresis of $\kappa(H)$ looks less
remarkable than the step-like features in the magnitude, it is
actually a more peculiar phenomenon related to the magnetism of
the spin-ice materials.

\subsection{Possible origins for the irreversibility of
$\kappa(H)$}

One possible origin of the irreversible $\kappa(H)$ behaviors is
related to field misalignment from the [111] direction, which is
known to have important impact on the field-induced
transitions.\cite{Tilting} When the field is slightly tilted from
the [111] direction due to a misalignment, its component parallel
to the (111) plane can break the threefold degeneracy of the
modified ice rule. Accordingly, the kagom\'e spins tend to align
at a temperature that depends on this parallel component. This is
known as a Kasteleyn transition.\cite{Tilting} An early entropy
measurement indicated that a field tilting larger than 2$^\circ$
can significantly smear out the local entropy peak at the
$H_{c2}$.\cite{Entropy} If such a misalignment of spins could
result in a hysteresis in the entropy of the kagom\'e-ice state
when the field is swept up and down, the phonon heat transport
would show a hysteresis considering the magnetic scattering on
phonons. Furthermore, the partial ordering caused by the parallel
component of field is likely accompanied with hysteresis because
of the slow spin dynamics. Since this kind of hysteresis may
appear only in the magnetization component perpendicular to the
[111] direction, it cannot directly be observed in the
magnetization along the [111] direction. Therefore, it would not
be surprising that the $\kappa(H)$ irreversibility has no direct
correspondence to the well-known $M(H)$ curves. For these reasons,
the field misalignment should be carefully taken into account.
However, there are two factors to be considered if one tries to
ascribe the irreversibility of $\kappa(H)$ to the field
misalignment. First, since both the low-$T$
$\kappa_{\parallel}(H)$ and $\kappa_{\perp}(H)$ isotherms in the
present work show rather sharp decreases at $H_{c2}$, the
misalignment in our measurements might be negligibly small.
Second, it has been not yet proved that such kind of field
misalignment can really lead to some irreversible behavior in the
kagom\'e-ice state.

\begin{figure}
\includegraphics[clip,width=8.5cm]{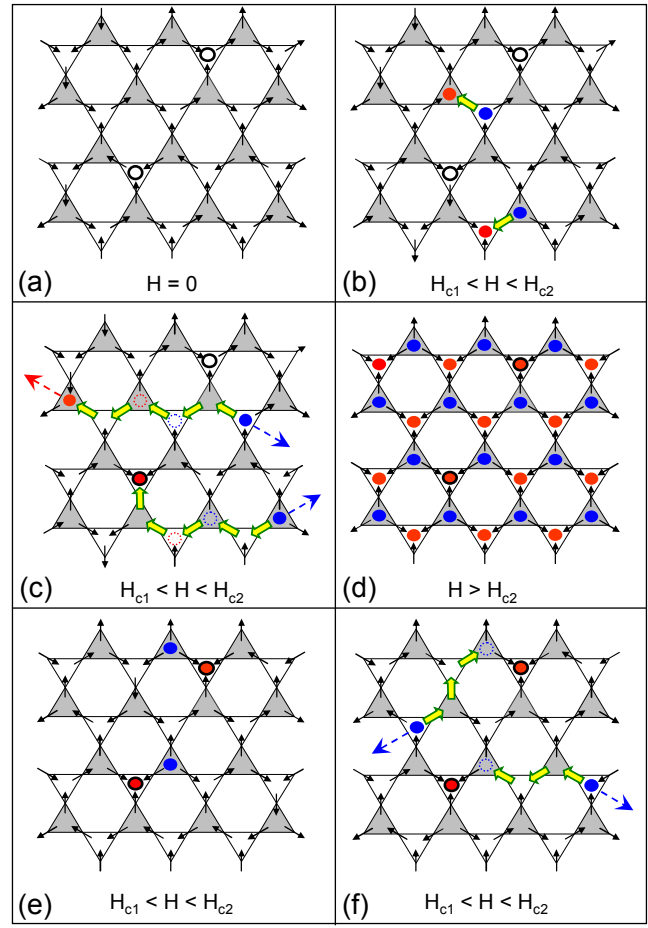}
\caption{(Color online) Structure evolution of the spin projection
on the (111) plane. The magnetic field is perpendicular to the
plane. $H_{c1}$ and $H_{c2}$ are defined in the text. Panels
(a)-(d) show the field-up process while (e)-(f) show the
field-down process. (a) In zero field and at very low
temperatures, the thermally excited magnetic monopoles are very
few. The black circles stand for crystal defects. (b) With
increasing field, some magnetic monopoles are created in pairs
(the red and blue dots). (c) Magnetic monopoles can move
separately by spin flipping. The flipped spins are shown by the
open and yellow arrows. The dashed red and blue arrows,
accompanied with the dashed open circles, show the routines of
positive and negative monopoles. Once some magnetic monopoles meet
with the defects, they can be pinned. (d) Above $H_{c2}$, the
magnetic monopoles are so well-populated that they cannot move any
more. (e) With decreasing field, the number of magnetic monopoles
is decreased. (f) Those magnetic monopoles sitting on the defect
sites are pinned, while the others still can move in the plane.}
\end{figure}

We propose another possibility with a ``pinning" effect of the
magnetic monopoles by the static disorders, in analogy to the
weak-disorder induced pinning effect on vortices in the
superconductors. Considering a simple case that the defect is on
the nonmagnetic Ti$^{4+}$ site, the interactions of the Dy$^{3+}$
spins in that tetrahedron would not be the same as others and the
flipping of these spins could be more difficult. In other words, a
magnetic monopole located at this defect site cannot move away
freely. Thus, crystal defects can work as pinning centers of the
magnetic monopoles. Figure 5 illustrates the movement process of
monopoles considering this kind of pinning effect. In zero field,
there are few magnetic monopoles induced by thermal fluctuations
(at very low temperatures), as shown in Fig. 5(a). When the field
is increasing, some spins are flipped and the magnetic monopoles
are created in pairs (see Fig. 5(b)) and diffuse freely in the
crystal (see Fig. 5(c)). With increasing the field further, the
density of magnetic monopoles increases and they could contribute
to scattering phonons when they are moving in the (111) plane;
this results in a quick suppression of $\kappa$, as shown in Figs.
4(a) and 4(b). Note that in this process, once some magnetic
monopoles meet with the defects, they can be pinned. Near the
saturation field, there are so many monopoles that they gradually
lose their mobility, and the scattering effect on phonons becomes
weakened and $\kappa$ can be somewhat recovered. At the saturated
state (the spins are fully polarized at $H_{c2}$), the magnetic
monopoles are well populated so that they fully occupy the
kagom\'e network with the monopole pairs sitting on the nearest
sites in Fig. 5(d). At this moment, they completely lose the
mobility and the $\kappa$ reaches the local maximum (the decrease
of $\kappa$ at $H > H_{c2}$ is due to other reasons, as was
discussed in section III. B). When the field is decreasing across
the saturation field, the density of magnetic monopoles is quickly
reduced by annihilation in pairs. Most of remained ones may have
some chance to start moving, but those pinned by the defects
cannot be mobile, as shown in Figs. 5(e)-(f). As a result, the
proportion of the mobile monopoles in the field-decreasing process
is smaller than that in the field-increasing case. Consequently,
the phonon scattering is weaker in the former case and $\kappa$
shows an irreversible behavior. It could be expected that the
irreversibility of $\kappa$ must be quickly weakened with
increasing temperature, because the stronger thermal fluctuations
will cause the magnetic monopoles to be depinned. One may think
that, based on the above discussion, the peculiar heat transport
properties can be explained by a simple phonon-assistant spin flip
and the magnetic monopoles might be irrelevant. However, it is
notable that the irreversibility is larger in
$\kappa_{\parallel}(H)$ than that in $\kappa_{\perp}(H)$. This
indicates that the magnetic monopoles can more effectively scatter
the in-plane phonons as a kind of quasi-particles since they can
move in the kagom\'e plane.\cite{monopole-3} The validity of this
picture needs further experimental and theoretical investigations.

\section{SUMMARY}

The very-low-temperature thermal conductivity of Dy$_2$Ti$_2$O$_7$
displays an irreversibility with magnetic field applied in the
[111] direction, which seems to have no direct correspondence with
the magnetization hysteresis. We discussed possible origins of
this irreversibility, including a field misalignment effect and a
pinning effect of magnetic monopoles by the weak disorders.

\begin{acknowledgements}

We thank Y. Takano for helpful discussions. This work was
supported by the National Natural Science Foundation of China, the
National Basic Research Program of China (Grant Nos. 2009CB929502
and 2011CBA00111), and the Fundamental Research Funds for the
Central Universities (Program No. WK2340000035).

\end{acknowledgements}

{\it Note added.}$-$We note a recent study on the heat transport
of Dy$_2$Ti$_2$O$_7$,\cite{Kolland} in which the measurements were
done in magnetic field along the [001] direction. An irreversible
behavior of $\kappa(H)$ was also presented in low fields ($<$ 0.3
T), which is directly related to the irreversible magnetization.
This is very different from what we observed in a much higher and
broader field range.

\end{document}